\documentclass[12pt]{article}
\usepackage{bm}
\usepackage{epsfig}
\usepackage{amssymb}
\usepackage{amsmath}
\usepackage{calrsfs}
\usepackage{multirow}
\usepackage{rotating}

\newcommand{\ep}{\varepsilon}

\newcommand{\bea}{\begin{eqnarray}}
\newcommand{\eea}{\end{eqnarray}}
\newcommand{\be}{\begin{equation}}
\newcommand{\ee}{\end{equation}}

\newcommand{\as}{\alpha_s}

\newcommand{\beq}{\begin{equation}}
\newcommand{\eeq}{\end{equation}}
\newcommand{\ar}{a_{\rm s}}

\newcommand{\eps}{\epsilon^{\:\!2}}

\def\fr#1#2{{\textstyle{{#1}\over{#2}}}}
\def\FR#1#2{{\displaystyle{{#1}\over{#2}}}}

\begin{document}

\setlength{\arraycolsep}{0.5mm}

\def\Qs{Q^{\:\! 2}}
\def\mus{\mu^{\:\! 2}}
\def\ca{{C_A}}
\def\cas{{C^{\: 2}_A}}
\def\cath{{C^{\: 3}_A}}
\def\cafo{{C^{\: 4}_A}}
\def\cafi{{C^{\: 5}_A}}
\def\cf{{C_F}}
\def\cfs{{C^{\: 2}_F}}
\def\cfth{{C^{\: 3}_F}}
\def\nf{{n^{}_{\! f}}}
\def\nfs{{n^{\,2}_{\! f}}}
\def\nft{{n^{\,3}_{\! f}}}
\def\nffo{{n^{\,4}_{\! f}}}

\title{
%About
SUSY, Casimir scaling  and probabilistic properties of gluon and quark jets evolution
%}
\author{
%Bernd~A.~Kniehl$^{1}$, Anatoly~V.~Kotikov$^{2}$\\
Anatoly~V.~Kotikov, Oleg~V.~Teryaev\\
Joint Institute for Nuclear Research, 141980 Dubna, Russia
}}

\maketitle

\begin{abstract}
  We study the new relation [B. A. Kniehl and A. V. Kotikov, arXiv:1702.03193.]
  %\cite{Kniehl:2017fix}
  between the anomalous
dimensions, resummed through next-to-next-to-leading-logarithmic order, in the
Dokshitzer-Gribov-Lipatov-Altarelli-Parisi
evolution equations for the first Mellin moments $D_{q,g}(\mu^2)$ of the fragmentation functions,
which correspond to the average
multiplicities of hadrons in jets initiated by quarks and gluons, respectively.
This relation is shown to lead to probabilistic properties of the properly rescaled parton
jet multiplicities obtained from standard ones by extracting the quark and gluon
"color charges" $C_F$ and $C_A$, respectively.  
\end{abstract}

%\pacs{12.38.Cy,12.39.St,13.66.Bc,13.87.Fh}

%\maketitle

\section{Introduction}

The broad and elegant concept of supersymmetry (SUSY) is currently manifested in various branches of physics.
For high energies it is pronounced in the properties of QCD supersymmetric extension rather than in the existence  of supersymmetric partners. In particular, this corresponds to the  SUSY-related properties of evolution kernels \cite{Gribov:1972ri} discovered some time ago \cite{Dokshitzer:1977sg}. In the current paper we explore the recently found relation \cite{Kniehl:2017fix} for fragmentation kernels and suggest 
its probabiolistic interpretation,  bringing SUSY closer to observations.

The notion of fragmentation functions (FFs)
$D_a(x,\mu^2)$  (hereafter $(a=q,g)$), where $\mu$ is the factorization scale,
was involved during the study of the inclusive
production of single hadrons.
Their $\mu^2$
dependence is
%are
governed by the timelike
%DGLAP
Dokshitzer-Gribov-Lipatov-Altarelli-Parisi (DGLAP) evolution
equations \cite{Gribov:1972ri,Dokshitzer:1977sg}
\begin{equation}
  \mu^2\frac{\partial D_a(x,\mu^2)}{\partial \mu^2}=\sum_{b} \, P_{ab}(x) \otimes D_b(x,\mu^2) \, ,
  \label{ap0}
\end{equation}
where $P_{ab}(x)$ are the time-like splitting functions and
the symbol $\otimes$ marks the Mellin convolution
\be
f_1(x) \otimes f_2(x) \equiv  \int_x^1 \frac{dz}{z} f_1\left(\frac{x}{z}\right) f_2(z) \, .
\label{MeConv}
\end{equation}

The DGLAP equations are conveniently solved in Mellin space (hereafter we represent the Mellin moment $N$ as $N=1+\omega$)
%($D_a(N,\mu^2)=\int dx\,x^{N-1}D_a(x,\mu^2)$
%%with
%$(N=1,2,\ldots)$
%are FF Mellin moments)
\begin{equation}
\mu^2\frac{\partial D_a(\omega,\mu^2)}{\partial \mu^2}=\sum_{b} \, P_{ab}(\omega)  D_b(\omega,\mu^2) \, ,
  %\frac{\mu^2d}{d\mu^2}
%\left(\begin{array}{l} D_s(N,\mu^2) \\ D_g(N,\mu^2) \end{array}\right)
%=\left(\begin{array}{ll} P_{qq}(N,a_s) & P_{gq}(N,a_s) \\
%P_{qg}(N,a_s) & P_{gg}(N,a_s) \end{array}\right)
%=\left(\begin{array}{ll} P_{qq}(N) & P_{gq}(N) \\
%P_{qg}(N) & P_{gg}(N) \end{array}\right)
%\left(\begin{array}{l} D_s(N,\mu^2) \\ D_g(N,\mu^2) \end{array}\right),
\label{apR}
\end{equation}
where
\be
D_a(\omega,\mu^2)=\int^1_0 dx\,x^{\omega}D_a(x,\mu^2)
\label{DaN}
\end{equation}
are FF Mellin moments. Here $P_{ab}(N,a_s)$ (hereafter $(a,b=q,g)$)
are anomalous dimensions (i.e. the Mellin moments of the corresponding splitting functions  $P_{ab}(x,a_s)$)
\be
P_{ab}(\omega,\mu^2)=\int^1_0 dx\,x^{\omega}P_{ab}(x,\mu^2)
\label{PabN}
\end{equation}
and
$D_s=(1/2n_f)\sum_{q=1}^{n_f}(D_q+D_{\bar{q}})$, with $n_f$ being the
number of active quark flavors, is the quark singlet component.
The quark non-singlet component
%, which
is irrelevant for the present study.

 The timelike splitting functions $P_{ab}(x,a_s)$ and the corresponding anomalous dimensions $P_{ab}(\omega,a_s)$ in
  Eq.~(\ref{apR})
  may be computed perturbatively in $a_s$,
\begin{equation}
 P_{ab}(x,a_s)=
\sum_{k=0}^\infty a_s^{k+1} P_{ab}^{(k)}(x),~~
  P_{ab}(\omega,a_s)=
\sum_{k=0}^\infty a_s^{k+1} P_{ab}^{(k)}(\omega) \, ,
\label{PijN}
\end{equation}
where 
  %and
  $a_s(\mu^2)=\alpha_s(\mu)/(4\pi)$ is the coupland.
The functions $P_{ab}^{(k)}(x)$ and $P_{ab}^{(k)}(\omega)$ for $k=0,1,2$ in the
$\overline{\mathrm{MS}}$ scheme may be found in
Refs.~\cite{Gluck:1992zx,Moch:2007tx,Almasy:2011eq} through the next-to-next-to-leading order (NNLO).
%and in 
%Refs.~\cite{Vogt:2011jv,Albino:2011si,Kom:2012hd} with small-$x$ resummation
%through NNLL accuracy.

The first Mellin moment $D_a(\mu^2)\equiv D_a(1,\mu^2)$ is of special
interest.
Up to corrections of orders beyond our
consideration here, this
%it
corresponds to the average
%hadron
multiplicity
$\langle n_h\rangle_a$ of hadrons in the jets initiated by parton $a$.
Now there are a lot
of experimental data on $\langle n_h\rangle_q$,
$\langle n_h\rangle_g$, and their ratio
$r=\langle n_h\rangle_g/\langle n_h\rangle_q$ for charged hadrons $h$ taken in
$e^+e^-$ annihilation at different
energies $\sqrt{s}$ of the center of mass, ranging
from 10 to 209~GeV  (see a list of references
in \cite{Bolzoni:2013rsa}).
The study of $D_a$ contains a long story:
the leading order (LO) value of $r$,
$C^{-1}=C_A/C_F$ with color factors $C_F=4/3$ and $C_A=3$, was found four
decades ago \cite{Brodsky:1976mg}.
\footnote{One should stress that the multiplicities $D_a(\mu^2)$ obey to so-called
    ``Casimir scaling'', since their results are given by universal function times
%    proportional to
the quadratic Casimir operators,
i.e. to $C_F$ and $C_A$ for the fundamental and adjoint representations \cite{Cvitanovic:1976am}
of the color SU(3) group, respectively
    (see Refs. \cite{Anzai:2010td}-\cite{Becher:2010tm} and discussions therein about the Casimir scaling, which
    %} {\it
    appeared in the 1980s \cite{Bernard:1982my} in lattice calculations).}

Usage of Eq. (\ref{apR}) with $N=1$ for
%The description of the $\mu^2$ dependences of
$D_a$ at fixed order in perturbation theory is problematic:
%are spoiled by the fact that
$P_{ab}\equiv P_{ab}(N=1)$ are
ill defined and require resummation, which was performed for the leading
logarithms (LL) \cite{Mueller:1981ex}, the next-to-leading logarithms (NLL)
\cite{Vogt:2011jv}, and the next-to-next-to-leading logarithms (NNLL)
\cite{Kom:2012hd}.

In Ref. \cite{Kniehl:2017fix} (see also \cite{Kniehl:2017oat}),
an unexpected relation between the NNLL-resummed expressions for $P_{ba}$ has been found.
%, which has gone unnoticed so far.
Its existence in QCD is quite remarkable and interesting in its own right,
because a similar relationship is familiar \cite{Dokshitzer:1977sg,Kom:2012hd,Kounnas:1982de}.
from supersymmetric QCD (SQCD),
where $C=1$.

In the present paper
  %  Here
  we will show that the relation obtained in Ref. \cite{Kniehl:2017fix}
  leads to probabilistic properties of the rescaled average multiplicities.

  The paper is organized as follows. In Section 2, we consider
  % show
  the basic properties and the results of the resummation
  for the anomalous dimensions of the fragmentation functions.
  The results are given by standart procedure \cite{Mueller:1981ex}, extended to $\overline{\mathrm{MS}}$ scheme in
  Ref. \cite{Albino:2011si}, as well by Vogt approach \cite{Vogt:2011jv,Kom:2012hd}.
  In Section 3, we present
  %  consider
  the two different procedures of diagonalization.
  In the first one, we  diagonalize the (LO of) the DGLAP equation for arbitrary $N$-values and later
  take the limit $\omega \to 0$ for $N=1+\omega$.
  Such diagonalization is also useful to study the full FF evolution.
  In the second possibility, we consider directly the first moment $N=1$
  and diagonalize the gluon and quark multiplicities themselves. Section 4 contains discussions of Casimir scaling and probabilistic
  properties of gluon and quark jet evolution.

\section{Resummation}

To explicate the ideas of resummation,
we consider here the cross section for the semi-inclusive hadron 
production in electron-positron annihilation:
\begin{equation}
e^+(k_1) + e^-(k_2) \rightarrow V^*(q)\rightarrow h(p_h)+X,
\label{epemg}
\end{equation}
where $V^*$ is a virtual vector boson with virtuality $Q^2=q^2=(k_1+k_2)^2$ and $X$
stands for any allowed hadronic final state.
Here we are interested in the differential cross section for the single hadron production
$d\sigma^h(x,Q^2)/dx$ (or the structure function $F(x,Q^2)=(d\sigma^h(x,Q^2)/dx)/\sigma^{(\rm ew)}_{k}$, where
$\sigma^{(EW)}$ contains all the electroweak over-all factors),
where $x$ is the scaled momentum fraction of the produced hadron $h$:
\begin{equation}
x=\frac{2p_h\cdot q}{Q^2},\qquad 0\leq x \leq 1.
\end{equation} 
According to QCD factorization the structure function $F(x,Q^2)$ can be written as 
a convolution of the coefficient function $C_a(x,Q^2)$ (i.e. the parton cross section to produce a parton $a$,
divided by factor $\sigma^{(EW)}$)
%with scaled momentum fraction $z=x/z'$
with the fragmentation function
$D_{a}(x)$ from the parton: 
\begin{equation}
  F(x,Q^2)
  %\equiv \frac{1}{\sigma^{(\rm ew)}_{k}}\, \frac{d\sigma^h}{dx}(x,Q^2)
  =\sum_a C_a(x,Q^2) \otimes D_{a}(x)\, ,
%  \int_x^1 \frac{dz'}{z'}\frac{d\hat{\sigma}_{e^+e^-\rightarrow k}}{dz} D_{k/h}(z'). 
\label{dsigma.1}
\end{equation}
where the symbol $\otimes$ defined in Eq. (\ref{MeConv}).
%marks the Mellin convolution
%\be
%f_1(x) \otimes f_2(x) \equiv  \int_x^1 \frac{dz}{z} f_1\left(\frac{x}{z}\right) f_2(z) \, .
%\label{MeConv}
%\end{equation}

As we already discussed in the introduction, perturbation theory does not work properly when the fraction $x$ of
available energy carried away by the observed particle is too small, since large logarithms spoil the convergence
of the perturbation theory series.
The largest logarithms, {\it double logarithms} (DLs) contribute to the splitting functions
$P_{a, b}(x,a_s)$, which determine the evolution of the fragmentation function, were computed for 
all orders long ago \cite {Mueller:1981ex}.
The total DL contribution to the parton cross sections was calculated in the work \cite{Mueller:1982cq} for the
case where collinear singularities are regularized by imparting a small mass $m_g$ to the  gluon, in the
so-called massive gluon (MG) regularization scheme.
As noted in  \cite{Albino:2011si} pioneering the NNLO approximation, the results found in
Ref. \cite{Mueller:1982cq} do not match the results computed at a fixed order (see Refs.
\cite{Rijken:1996vr,Mitov:2006ic,Blumlein:2006rr}) whoch is not surprising, because the two computations were
performed in two different regularization and factorization schemes, namely MG and the $\overline{\rm MS}$ scheme.
DL gluon coefficients were calculated in the $\overline{\rm MS}$ scheme in \cite{Albino:2011si}.
%Here
In the subsection 2.1 we present the result in a less formal but simpler way.

%Here we
So, below we demonstrate the basic ideas of the resummation
of the fragmentation functions,
which is important near its first Mellin
moment. We consider an extension to the $\overline{MS}$-scheme \cite{Albino:2011si} (see also a short review in Ref.
\cite{Albino:2011bf})
of the traditional approach
\cite{Mueller:1981ex,Mueller:1982cq}
of resummation (see Section 2.1)
and also show the important points of the Vogt et al. approach  \cite{Vogt:2011jv,Kom:2012hd}, which gives a possibility to have
an accurate treetment of the resummation upto
NNLL level of accuracy.

\subsection{Traditional approach to resummation}

As it  is well known, these DL contributions appear in the gluon-gluon and gluon-quark
timelike splitting functions \cite{Mueller:1981ex} and in the timelike gluon coefficient function
\cite{Mueller:1982cq}.
To extract them, consider a general process with a final singlet color state, including the creation of
an "observed" gluon with momentum $q$ from a hard parton with momentum $p$ around which a jet is formed.
In the DLA, the DL contributions arise from unobservable soft gluons in the final state.
Therefore, there must be an additional hard parton to take into account  the recoil from the parton of momentum $p$
 as a result of momentum conservation. The momentum of this extra parton is written as $\bar{p}$.
The cross section for this process will be written $d\sigma(p,\bar{p},q)$.
A typical example is the process
$e^+ + e^- \rightarrow V^* \rightarrow {\mathcal Q}(p) +\bar{\mathcal Q}(\bar{p}) +g(q) +X $, 
where $V^*=\gamma, Z$ - virtual vector boson, where the jet is formed around the quark ${\mathcal Q}$ with
momentum $p$ and around the antiquark $\bar{\mathcal Q}$ with momentum $\bar{p}$, and where $X$ is any
hadronic final state allowed by the conservation of the quantum number.
Thus, in order to obtain the DL contribution  to the cross section, we consider a configuration in which the
unobservable part consists only of $N$ soft gluons of momenta $q_1,q_2,\ldots,q_N$, whose phase space is
fully integrated.
Therefore, defining $d\sigma_N(p,\bar{p},q_1,q_2,\ldots,q_N)$ as a section in which $N$ gluon momenta
$q_\alpha$, $\alpha=1,2,\dots,N$, are produced together with $p$ and $\bar{p}$ momentum partons, we can write
\be
d\sigma(p,\bar{p},q)=\sum_{N=0}^\infty d\sigma_{N+1}(p,\bar{p},q,q_1,q_2,\ldots,q_N),
\label{dsigppprimeq}
\ee
where it is understood that the $q_\alpha$ are fully integrated over, but not $q$.
It is a well known result \cite{Bassetto:1982ma} that the DL
contributions come from the kinematic configuration in which the momenta
of the soft gluons and also
the angles $\theta_i$ of the emitted
gluons with respect to the hard parton of momentum $p$
are strongly ordered, i.e.
\be
|\vec{q}|\ll |\vec{q_1}|\ll |\vec{q_2}|\ll \cdots \ll |\vec{q_N}| \ll Q/2,~~
\theta \ll \theta_1\ll\theta_2\ll\cdots\ll\theta_N\ll 1,
\label{momord}
\ee 
where  $\theta$ refers to the gluon of momentum $q$.
%and $Q^2$ is the perturbative scale, i.e. \
%the scale that the factorization and renormalization scales should have the same
%order of magnitude as. For a general process, the choice $Q^2 =(p+\bar{p})^2$ is suitable and is usually made, in particular this choice is made in $e^+ e^-$ annihilation. 

To extract the LL
%leading logarithmic
behavior, we use the single-gluon probability emission
factorization in the soft collinear limit.
This is a consequence of the eikonal approximation
and  color coherence, as has been proven long ago in \cite{Bassetto:1982ma,Bassetto:1984ik}.
This result was used \cite{Albino:2011si} to obtain the probability of gluon emission in $d=4-2\epsilon$ dimensions:
\begin{equation}
  dw(x,z,\epsilon)=2 C_i
  %\mathcal{C}_i
  \, a_s \left(\frac{\mu^2}{Q^2}\right)^\epsilon \frac{(4\pi)^\epsilon}{\Gamma(1-\epsilon)} 
\frac{dx}{x^{1+2\epsilon}}\, \frac{dz}{z^{1+\epsilon}},
\label{probemission}
\end{equation} 
%where $a_s=\alpha_s/2 \pi$, $\mu$ is the dimensional regularization scale taken here and in the following
%equal to the renormalization scale and
where $z=(1-\cos\theta)/2$ with $\theta$ the scattering angles of the emitted soft gluon with respect to
the hard jet direction. Here $C_i=C_A$ for a gluon jet and $C_i=C_F$ for a quark jet.
The expression for the probability emission given in Eq.(\ref{probemission}) is 
what we need to obtain the gluon probability density in dimensional regularization. 
%The expression for probability emission given in  Eq.(\ref{probemission}) is what we need to get the gluon
%probability density in dimensional regularization.

The consistency relation for the differential cross section for the gluon jet production
has the following form
\begin{equation}
d\sigma_g^n=d\sigma_g^1+d\sigma_g^{n-1}dw(x,z,\epsilon).
\end{equation}
%Now introducing $G(x,\epsilon)$ the gluon distribution density and
Taking the limit $n\rightarrow \infty$, we obtain immediately
the following bootstrap equation
\begin{equation}
x^{1+2\epsilon}\mathcal{G}(x,z,\epsilon)=
\delta(1-x)+\int_x^1 dx' \int_z^1 dz'\,K(x',z',\epsilon)\, x'^{1+2\epsilon} \mathcal{G}(x',z',\epsilon), 
\label{mastereq}
\end{equation}
where $K(x,z,\epsilon)=dw(x,z,\epsilon)/dx\,dz$ and $\mathcal{G}(x,z,\epsilon)$ is an angle-dependent gluon density.
%where the
The factor of 
$x^{1+2\epsilon}$ represents our normalization
coming from the explicit computation for the first gluon emission with $n=2$.

Now introduce  the gluon distribution density $G(x,\epsilon)$ as
%In Eq.(\ref{mastereq}) the value
$\mathcal{G}(x,z=0,\epsilon)=G(x,\epsilon)$.
%should be taken only at the end of the computation.
%This ensures the strong angular
%ordering of the emitted gluons, which is necessary to extract correctly 
%the leading logarithms as proven in Refs.\cite{Mueller:1981ex,Bassetto:1982ma}.
Let us perform  the Mellin transform, of Eq.(\ref{mastereq}) as well as in Eqs.  (\ref{DaN}) and (\ref{PabN})  above.
%\begin{equation}
%f(\omega)=\int_0^1\, dx x^\omega f(x); \qquad \omega=N-1,
%\end{equation}
Further, integrating over $z$,   
solving recursively  the equation for $\mathcal{G}(x,z,\epsilon)$ and finally putting $z=0$
we obtain (see details in Ref. \cite{Albino:2011si})
%we can rewrite 
\be
  G(\omega,\epsilon)=1+\sum_{k=1}^\infty
  %\left[2a_sC_A\frac{(4\pi)^\epsilon}{\Gamma(1-\epsilon)}
  %\left(\frac{\mu^2}{Q^2}\right)^\epsilon\right]^k
  \left(\frac{X}{2\ep^2}\right)^k \, \frac{\Gamma(1+\nu)}{k!\Gamma(k+1+\nu)}
  %\frac{\Gamma(1-\omega/(2\epsilon)}\frac{k!\Gamma(k+1-\omega/(2\epsilon)}
= \Gamma(1+\nu) \, \left(\frac{Z}{2}\right)^{\nu} \, I_{\nu}(Z) \,
\label{solution.1}
\end{equation}
where
\be
X=2C_A a_s \frac{(4\pi)^\epsilon}{\Gamma(1-\epsilon)}\left(\frac{\mu^2}{Q^2}\right)^\epsilon = 2C_A a_s + O(\ep)
\label{X}
\ee
%where
and $\nu=-\omega/(2\epsilon)$ and $z=\sqrt{2A}/\ep$ and $I_{\nu}(z)$ is modified Bessel function.

According to the QCD factorization theorem, all collinear singularities in 
Eq.(\ref{solution.1}) should be factorized. In the $\overline{\rm MS}$ factorization
scheme, this requires comparing  Eq.(\ref{solution.1}) with (see \cite{Curci:1980uw,Catani:1994sq})
\begin{equation}
G(\omega,\epsilon)=C^{\overline{\rm MS}}\left(\omega,a_s,\frac{Q^2}{\mu_F^2}\right)
\exp\left[-\frac{1}{\epsilon} \int_0^{a_s(\mu_F^2)}
  %\int_0^{a_s(\mu^2/\mu_F^2)^\epsilon S_\epsilon}
  \frac{da}{a}
P^{\overline{\rm MS}}(\omega,a)\right],
\label{factorization}
\end{equation}
to the proper accuracy.
%where $S_\epsilon=(4\pi)^\epsilon e^{-\epsilon\gamma_E}$ with $\gamma_E$ the Euler number
%and where $\mu_F$ is the arbitrary factorization scale.

The direct comparison of equations (\ref{solution.1}) and (\ref{factorization}) is highly non-trivial. However,
it is well-known that Bessel function obey the second order differential equations. 
%Now, direct factorization of the collinear singularities as $\epsilon \rightarrow 0$ in Eq.\ (\ref{finformofG}) by expanding it in $\epsilon$ does not seem possible. 
%However
Indeed, from Eq.\ (\ref{solution.1}), it is easy to check that $G$ satisfies the following simple
differential equation:
\be
\frac{\ddot{G}-\frac{\omega}{2\epsilon}\dot{G}}{G}=\frac{X}{2\epsilon^2},~~ \dot{f}(X) \equiv X\frac{d f}{dX} = a_s\frac{d f}{da_s} \, .
\label{diffeq1}
\ee
%where we have defined
%\be
%\dot{f}(X)=X\frac{d f}{dX}.
%\label{devdef}
%\ee

%Now we can study the Eq. (\ref{factorization}). For simplicity, we will set $\mu_R=\mu_F=Q$ from now on, so that
%\be
%X=2C_A  a_s(Q^2,\epsilon)
%\ee
%for DL accuracy.
From Eq. (\ref{factorization}), we have
\be
\frac{\dot{G}}{G}=%\left(
\frac{\dot{C}}{C}-\frac{P}{\epsilon},~~
%C\right)\Gamma,~~
\frac{\ddot{G}}{G}=
%\left(
\frac{\ddot{C}}{C}-2\frac{P}{\epsilon}\frac{\dot{C}}{C}-\frac{\dot{P}}{\epsilon}+\frac{P^2}{\epsilon^2} \, ,
%C\right)\Gamma,
\ee
so that
\be
\frac{\ddot{G}-\frac{\omega}{2\epsilon}\dot{G}}{G}=\frac{1}{\epsilon^2}\left(P^2+\frac{\omega P}{2}\right)
-\frac{1}{\epsilon}\left(\left[2\gamma +\frac{\omega}{2}\right]\frac{\dot{C}}{C}+\dot{P}\right)+\frac{\ddot{C}}{C}.
\label{diffeqforGCandgamma}
\ee
%Note that keeping the factor $K_\epsilon=1+O(\epsilon^2)$ in Eq.\ (\ref{Xfromas}) 
%would result in an $O(\epsilon^0)$ change in this expression coming
%from the first term of $O(\epsilon^{-2})$, thus $K_\epsilon$ can be safely excluded.
Now, comparing the coefficients of $\epsilon^{-2}$ and $\epsilon^{-1}$ on the right 
hand side of Eq.\ (\ref{diffeqforGCandgamma}) with those of Eq.\ (\ref{diffeq1})
and noting that $\gamma$ is explicitly independent of $\epsilon$ one get, respectively,
\be
P^2+\frac{\omega\, P}{2}-\frac{X}{2}=0,~~
%\label{eqforgamma}
%\ee
%and
%\be
\frac{\partial \ln C}{\partial P}=-\frac{2}{4P+\omega}.
%\label{eqforC}
\label{eqforgamma}
\ee
%We report here the result which is given by:
%\begin{equation}
%  %G^{\overline{\rm MS}}
%  G\left(\omega,a_s,\frac{Q^2}{\mu_F^2}\right)=
%  %C^{\overline{\rm MS}}
%  C(\omega,a_s)
%  \exp\left[%\gamma^{\overline{\rm MS}}
%    \gamma(\omega,a_s)\log\left(\frac{Q^2}{\mu_F^2}\right)\right],
%\label{result}
%\end{equation}
%where frm equation (\ref{eqforgamma})
%and (\ref{eqforC})
and, thus, we have
%, respectively,
\begin{equation}
  P(\omega,a_s)=\frac{1}{4}\left[\sqrt{\omega^2+32C_Aa_s]}-\omega\right]=
    \frac{1}{4}\left[\sqrt{\omega^2+16\gamma_0^2}-\omega\right],~~ 
C(\omega,a_s)=\left[\frac{\omega}{4P(\omega,a_s)
+\omega}\right]^{\frac{1}{2}}
\label{gamma+c}
\end{equation}
%and 
with
\be
\gamma_0=\sqrt{2C_Aa_s} \, .
\label{gamma0}
 \ee
%Expanding the result in Eq.(\ref{result}) up to NNLO we obtain perfect agreement with the leading logarithmic
%terms of the fixed order result computed in the literature in the same scheme (see e.g. Eqs.(A.3) and (A.6)
%in Ref.\cite{Blumlein:2006rr}).

 So, after the resummation, we have a flat limit at $\omega \to 0$ and
 \be
 P(\omega=0,a_s)= \gamma_0,~~ C(\omega,a_s) = \left[\frac{\omega}{4\gamma_0
+\omega}\right]^{\frac{1}{2}}
 %+ O(\gamma_0^2) \, ,
 \label{gamma0}
 \ee
%where the term $\sim  O(\gamma_0^2)$ contain the terms starting with $a_s$.
and $ \gamma_0$ appears as a new parameter of expansion.

\subsection{Vogt approach}

Here we show the basic properties
%elements
of A. Vogt approach
%most important st  Here thanks to the approach used by A.\,Vogt in Ref.
\cite{Vogt:2011jv}, where according to the all order factorization the singularities of the splitting functions
are iteratively extracted into the transition function $Z(\epsilon)$ 
of the $\epsilon=0$
poles in dimensional regularization ($d=4-2\epsilon$). 

%Indeed, according
According to the factorization theorem \cite{Ellis:1978ty,Curci:1980uw}, we can rewrite the expression (\ref{dsigma.1})  for the structure function $F(x,Q^2)$
 in Mellin space
\footnote{In $e^+e^-$ process there are several structure functions but we consider only $F_A$. A consideration of other
  functions  is similar and can be found in Ref. \cite{Vogt:2011jv}.}
as
%$F_m$ $(m=T,A,L)$
%%Eq.\ref{dsigma.1}, i.e.
%\begin{equation}
%F_m(x)=\sum_{a} C_{ma}(x)\otimes D_a(x),\label{factor}
%\end{equation}
%with the fragmentation function  $D_a(x)$, which obeys the DGLAP equation (\ref{ap0}).
%%\begin{equation}
%%Q^2\frac{\partial D_a(x,Q^2)}{\partial Q^2}=\sum_{b} \, P_{ab}(x) \otimes D_b(x,Q^2) \, \label{ap0}
%%\end{equation}
%In Mellin space Eq. (\ref{factor}) becomes to be
%as
\begin{equation}
  F(\omega)=\sum_{a} C_{a}(\omega)\,D_a(\omega)=\sum_{a,b,c} \hat{C}_{a}^0(\omega,\epsilon)Z_{ab}^{-1}(\omega,\epsilon)\,
  Z_{bc}(\omega,\epsilon) \hat{D}_{c}^0(\omega),\label{trans}
\end{equation} 
where $\hat{C}_{a}^0$ and $\hat{D}_{c}^0$ are the so-called ``bare'' coefficient function and fragmentation function.
$Z_{ab}(\epsilon)$ is the transition function containing only poles in $\epsilon$ that are factored out
from $\hat{C}_{a}^0$.
Hence, substituting Eq.(\ref{trans}) into Eq.(\ref{ap0}) at $\mu^2=Q^2$ we find that
the splitting functions can be directly related to the transition function in the following way:
\begin{equation}
  P_{ab}(\omega)=\sum_{c} Q^2\frac{\partial Z_{ac}(\omega)}{\partial Q^2} Z_{cd}^{-1}(\omega)=\beta_{\rm D}(\alpha_s)\sum_{c} 
  \frac{\partial Z_{ac}(\omega)}{\partial \alpha_s}Z_{cb}^{-1}(\omega) \, ,
\label{Factor}
\end{equation}
where
\be
\beta_{\rm D}(\alpha_s)= \ep a_s - \sum_{i=0} \beta_i a_s^{i+1} \, 
\label{betaD}
\ee
and $\beta_0$ and $\beta_1$ values are given below in Eq. (\ref{eq:beta}).

As it was shown in Ref.\cite{Vogt:2011jv},
%how
one can solve the equation (\ref{Factor}) in $Z_{ab}$ obtaining
at all orders the three highest order poles in $\epsilon$ knowing the NNLO corrections to $P_{ab}$ and $\beta_i$. 
Additionally knowing the higher order corrections to $C_{a}$ which is pole free one 
obtains from the NNLO computations the all order structure of the three first highest singularities in 
$\epsilon$:
\begin{equation}
 \hat{C}^0_{a}(\omega,\epsilon)=\sum_{b} C_{b}(\omega,\epsilon)Z_{ba}(\omega,\epsilon).\label{epspoles}
\end{equation}
$\hat{C}^0_{a}(\omega,\epsilon)$ may be computed perturbatively in $a_s$
\be
 \hat{C}_{a}^0(\omega,\epsilon)= \sum_n \alpha_s^n \,  \hat{C}_{a}^{(n)}(\omega,\epsilon),~~
\Bigl( \mbox{and} ~
%\label{CN}
%\end{equation}
 \hat{C}_{a}^0(x,\epsilon)= \sum_n \alpha_s^n \,  \hat{C}_{a}^{(n)}(x,\epsilon) ~\mbox{in $x$-space} \Bigr) \, .
\label{Cx}
\end{equation}
%in $x$-space and
%\be
%C_{ma}^0(N,\epsilon)= \sum_n \alpha_s^n \, C_{ma}^{(n)}(N,\epsilon)
%\label{Cx}
%\end{equation}
%in momentum space.

The key point of Ref.\cite{Vogt:2011jv} is that , for example, for the case 
of the gluon
\footnote{The quark function $\hat{C}_{q}^{(n)}(x,\epsilon)$ can be analyzed similarly (see  Ref.\cite{Vogt:2011jv}).}
the small $\omega$ behavior of the bare coefficient functions $\hat{C}_{g}^{(n)}(x,\epsilon)$
%The new functions $C_{a}^{(n)}(x,\epsilon)$ has the expnasion
\beq
\label{FagDx}
  \hat{C}_{g}^{\,(n)}(x,\ep) \;\;=\;\;
  \FR{1}{\ep^{\,2\,n-1}}\; \sum_{\ell=0}^{n-1}
  \:\: x^{\,-1-2\:\!(n-\ell)\:\!\ep}
  \, \big( \, A_{g}^{\,(\ell,n)} \,+\, \ep\, B_{g}^{\,(\ell,n)}
  \,+\, \eps\, C_{g}^{\,(\ell,n)} \,+\:\ldots\, \big) \qquad 
\eeq
up to
%terms of order $\omega^{\,0}$, i.e.,
non-singular in $x$ contributions.
%Eqs.~(\ref{FagD}) and (\ref{FagDx}) and the corresponding results for 
%$\widehat{F}_{T, \rm q}$, $\widehat{F}_{\phi, \rm q}$ and $\widehat{F}_{L,i}$ 
%given below form the crucial observation of this article. 
 
Focusing for a moment on the leading logarithms, Eq.~(\ref{FagDx}) provides decomposition of 
$\hat{C}_{g}^{\,(n)}$, which includes terms of the form 
$x^{\,-1} \ln^{\,n+m-1\!} x\,$ at all orders $\ep^{\,-n+m}$ with 
$m = 0,\,1,\,2,\:\ldots\,$, into $\,n$ contributions of the form
\beq
\label{xtoep}
  \ep^{\,-2n+1}\, x^{\,-1-k\,\ep} \;\;=\;\; \ep^{\,-2n+1}\, x^{\,-1}
  \left[ 1 \,-\, k\,\ep \ln x \,+\, \fr12 (k\,\ep)^2 \ln^{\,2\!} x \,+\: 
  \ldots \,\right]
\eeq
with $k = 2,\,4,\, \ldots,\, 2n$. Since $\hat{C}_{g}^{\,(n)}$ 
starts only at the order $\ep^{\,-n}$, the coefficients $A_{g}^{\,(\ell,n)}$ in 
Eq.~(\ref{FagDx}) have to be such that the coefficients of $\,\ep^{\,0},\, 
\ldots\,,\, \ep^{\,n-2}$ in the square bracket in Eq.~(\ref{xtoep}) cancel in 
the sum of these $n$ contributions. Together with the three non-vanishing 
coefficients of $\ep^{\,-n+\ell}$, $ \ell = 0,\,1,\,2\,$, in 
$\hat{C}_{g}^{\,(n)}$ known from the above NNLO results, author of \cite{Vogt:2011jv}  thus 
had an overconstrained system of $n+2$ linear equations for the $n$ 
coefficients $A_{g}^{\,(\ell,n)}$ at each order $n$ of the strong 
coupling. Similar results have been obtained also for the coefficients $B_{g}^{\,(\ell,n)}$ and $C_{g}^{\,(\ell,n)}$.
%It is non-trivial that all these systems have solutions, e.g., there would be 
%no solutions if the factor of two in front of $(n-\ell)$ in Eqs.~(\ref{FagD}) 
%and (\ref{FagDx}) was absent, or if the sign of this term was different.

In the Mellin space Eq. (\ref{FagDx}) transforms to
\begin{equation}
  \hat{C}_{g}^{(n)}(\omega,\epsilon)= \,  \frac{1}{\epsilon^{2n-1}}
  %\frac{1}{\omega}\,  \sum_n \frac{\alpha_s^n}{\epsilon^{2n-1}}
  \sum_{l=0}^{n-1} \frac{1}{\omega-2(n-l)\epsilon}
  %\frac{1}{1-2(n-l)\epsilon/\omega}
  (A_{g}^{(l,n)}+\epsilon B_{g}^{(l,n)}+\epsilon^2 C_{g}^{(l,n)}+\dots).\label{vogtguess}
\end{equation}
Finally comparing Eq.(\ref{vogtguess}) with Eq.(\ref{epspoles}) one gets the systems of equations for 
the coefficients $A_{g}$, $B_{g}$ and $C_{g}$, which produce sequences up to arbitrary orders in $\alpha_s$
of the three highest powers in $1/\omega$ or equivalently (back to $x$-space) in $\ln x$. 
%Then the highly non trivial
%part of this approach is the solution of these sequences that are obtained for the large logarithms
%and this has been successfully obtained in Refs.\cite{Vogt:2011jv,Kom:2012hd}.
%%The LL resummation leads exactly to results (\ref{gamma+c}).

Here
%In this section
we present the resummed timelike LL and NLL splitting functions  obtained in Refs.\cite{Vogt:2011jv,Kom:2012hd}
%to next-to-next-to-leading logarithmic (NNLL) accuracy,
%
\beq
\label{Pggexp}
  P_{ab}(\omega) \;\:=\;\: \sum_{n=0}^{\infty}\, \ar^{\,n+1} 
  \left(  \delta_{\rm \,ag\,}^{} P_{\,\rm ab,\,LL}(\omega)
   \:+\: P_{\,\rm ab,\,NLL}(\omega)
 %  \:+\: P_{\,\rm ab,\,NNL}(\omega) 
   \:+\; \ldots \right) \; ,
\eeq
where $ \delta_{\rm \,ag\,}$ is Kronecker symbol.

The LL and NLL
%leading log (LL) and next-to-leading log (NLL) 
contributions for $P_{\rm gg}$
%and $P_{\rm gq}$
have the form
\bea
\label{PggLL}
   &&P_{\rm gg,\,LL}(\omega) \;\:=\;\: 
  % \frac{\ca}{\cf}\: P_{\rm gq,\,LL}(\omega) \;\:=\;\:
   - \; \frac{(-8\,C_A)^{n+1}}{2\omega^{2n+1}}\: A^{(n)}_{\rm gg} ; \\
%\eeq
%and
%\bea
%\label{PggNL}
  &&P_{\rm gg,\,NLL}^{\,(n)T}(N) \!=\!
   - \; \frac{(-8)^{\,n}\,C_A^{\,n-1}}{3\omega^{2n}} \:
   \Big[ (11\,\cas + 2\,\ca \nf ) \, B^{\,(n)}_{\rm gg,1}
   \,-\, 2\: \cf \nf\, B^{\,(n)}_{\rm gg,2} \Big] \; .
\eea
The coefficients in Eq.~(\ref{PggLL})
%-- (\ref{PgqNL})
have been determined
to order $\as^{\,16}$, that leads to the following analytic results:
%($n=15$ in Eq.~(\ref{Pggexp})).\\
%, and are given in Table 
%\ref{tab:pginl} to the tenth order in $\as$ -- for the next six orders see 
%the text below Eq.~(\ref{Aqifrom10}).\\
%The general form and generating function for these series are known at this 
%point (to this author) only for Eq.~(\ref{PggLL}) and the non-$C_F$ terms in 
%the square brackets in Eqs.~(\ref{PggNL}) and (\ref{PgqNL}), i.e., those 
%entries that do not involve factorial denominators. 
%$A^{(n)}_{\rm gi}$ are the Catalan numbers
%\cite{OEIS,Catalan},
%
\beq
\label{Agi}
  A^{(n)}_{\rm gg} \;=\; \frac{ (2n)! }{ n! (n+1)! } \;=\; {1 \over n+1}
  \left( \begin{array}{cc} \!\! 2\:\!n \!\! \\ n \end{array} \right) \; ;~~
%\eeq
%and $B^{\,(n)}_{\rm gg,1}$
%%and $B^{\,(n)}_{\rm gq,2}$
%are given as
%%by \cite{binom12}
%%
%\beq
%\label{Bgg1}
  B^{\,(n)}_{\rm gg,1} \;=\; \left( \begin{array}{cc} \!\! 2\:\!n-1 \!\!
                             \\ n \end{array} \right) 
  \;.
  %\quad
  %B^{\,(n)}_{\rm gq,2} \;=\; \left( \begin{array}{cc} \!\! 2\:\!n-1 \!\! 
   %                          \\ n-2 \end{array} \right) 
  %\;=\; B^{\,(n)}_{\rm gg,1} \,-\, A^{(n)}_{\rm gi} \; ,
\eeq

The results (\ref{Agi})
%and (\ref{Bgg1})
lead to the closed NLL expressions
\be
\label{Pgg-cl}
  P_{\,\rm gg}(\omega) \Big|_{C_F=0} \!\!=\!  \; \fr{1}{4} \:
  \left[ \! \sqrt{\omega^2 + 16\gamma_0^2}\! - \omega \right] 
  %\nn \\[1mm] & & \mbox{\hspn}
  - \: \ar \, \Big( \,\fr{11}{6}\:\ca + \fr{1}{3}\:\nf \Big)
\left[ \frac{\omega}{\sqrt{\omega^2 + 16\gamma_0^2}}
%\left( 1 -4\,\xi \right)^{-1/2}\!
+ 1 \right]
  \   
  \;+\; ... 
  %  P_{\,\rm gg,\,NNL}(\omega)
  \; , 
\ee
with $\gamma_0$ is defined in eq. (\ref{gamma0}).

We see the full agreement of the LL part with  $P(\omega,a_s)$ in Eq. (\ref{gamma+c}). To obtain more complicated NNLL results
authors of \cite{Vogt:2011jv,Albino:2011bf} used the results (\ref{Agi}) as an initial form to present an ansatz for the
corresponding NNLL results (see \cite{Vogt:2012gb} and discussions therein).
\footnote{Similar ideas have been used also to find \cite{Fleischer:1998nb} an ansatz for  $\ep$-expansion coefficients in
  (series representations) of massive diagrams and also to generate \cite{Kotikov:2007cy,Kotikov:2008pv}
  an ansatz in calculations in the framework of ${\mathcal N}=4$ super Yang-Mills (SYM) theory.
  %  $N=4$ SYM.
  In particular, such approach gives a possibility to find a so-called
  universal anomalous dimension in ${\mathcal N}=4$ SYM up to seven loops (see \cite{Marboe:2016igj} and references therein).}

\subsection{Results for multiplicities}

The NLL and NNLL results for $P_{ab}(\omega)$ are rather cumbersome and can be found in Refs.
  \cite{Vogt:2011jv,Albino:2011si}.
So, here we present only the results for multiplicities, i.e. for $P_{ab}(\omega=0)\equiv P_{ab}$.

Consider now
%Our starting point is
Eq.~(\ref{apR}) for $\omega=0$
%$N=1$
with NNLL resummation, where
%We have
\cite{Kom:2012hd}
\begin{eqnarray}
P_{aa}&=&\gamma_0(\delta_{ag} + K_{a}^{(1)} \gamma_0
+ K_{a}^{(2)}  \gamma_0^2) ,~~
P_{gq}=
%&=&
C (P_{gg} +A),~~
P_{qg} =
%&=&
C^{-1} (P_{qq} +A) ,
%+ \mathcal{O}(\gamma_0^4),
\label{NNLL}
\end{eqnarray}
with $\mathcal{O}(\gamma_0^3)$ accuracy,
where $\gamma_0$ is given in Eq. (\ref{gamma0}),
%$\gamma_0=\sqrt{2C_Aa_s}$,
%%with
%$a_s=\alpha_s/(4\pi)$ is
%%being
%the couplant,
$\delta_{ab}$ is the Kronecker symbol, and
\begin{eqnarray}
K_{q}^{(1)} &=& \frac{2}{3} C\varphi,\quad K_{g}^{(1)} =
%&=&
-\frac{1}{12}[11 +2\varphi (1+6C)],\quad
K_{q}^{(2)} = -\frac{1}{6} C\varphi [17-2\varphi(1-2C)],
\nonumber \\
%K_{g}^{(1)} &=& -\frac{1}{12}[11 +2\varphi (1+6C)],\quad
K_{g}^{(2)} &=& \frac{1193}{288} -2\zeta(2)
%\nonumber \\&&{}
- \frac{5\varphi}{72}(7-38C)+\frac{\varphi^2}{72}(1-2C)(1-18C),~
%\nonumber \\
A =
%&=&
K_{q}^{(1)}\gamma_0^2 \, .
%\varphi=\frac{n_f}{C_A}.
\label{nllfirstA}
\end{eqnarray}
with $C=C_F/C_A$ introduced above and
\be
\varphi=\frac{n_f}{C_A} \, .
\label{vp}
\ee
Eq.~(\ref{NNLL}) is written in a form that allows us to glean a novel
relation (see \cite{Kniehl:2017fix}):
\begin{equation}
P_{qq}+ C^{-1}P_{gq} = P_{gg} + CP_{qg},
\label{Basic}
\end{equation}
which is independent of $n_f$.

Note that the form of the equation (\ref{NNLL}), as well as the equation (\ref{Basic}) itself, was obtained  in \cite{Kniehl:2017fix} by
  diagonalizing the quark and gluon multiplicities (see Subsection 3.1 below). Quite unexpectedly,
  in the results for diagonal anomalous dimensions (\ref{Ppm}), all square roots were cancelled and the equations (\ref{alpha1}) were obtained.
  Such root cancellation was previously observed only in supersymmetric generalizations of QCD. The reason for this simplification is associated with
  the resummation of the anomalous differences and with the direct diagonalization of parton multiplicities done in Subsection 3.1, which leads to some kind
  of symmetry. It is very important to look at its possible violation in higher orders of the perturbation theory and in the resummation
of higher logarithms. This is a very difficult problem requiring further investgation.

\section{Diagonalization}

%\subsection{Diagonalization}

In the general case, it is impossible to diagonalize Eq.~(\ref{apR}), since the contributions to the matrix of the
timelike splitting functions do not commute in different orders.
The usual approach is to write a series expansion for the LO solution, which in turn can be diagonalized.
Thus, we start by choosing a basis in which the matrix of the timelike LO splitting function is diagonal
(see, e.g., Ref.~\cite{Buras:1979yt}),
%{\bf
%  \begin{equation}
\bea
&&U_\omega^{-1}\left(\begin{array}{ll}
P^{(0)}_{qq}(\omega) & P^{(0)}_{gq}(\omega) \\ P^{(0)}_{qg}(\omega) & P^{(0)}_{gg}(\omega)
\end{array}\right)U_\omega
=\left(\begin{array}{ll}
P^{(0)}_{--}(\omega) & 0 \\ 0 & P^{(0)}_{++}(\omega)
\end{array}\right),~~ \nonumber \\
&&U_\omega^{-1}\left(\begin{array}{ll}
P^{(k)}_{qq}(\omega) & P^{(k)}_{gq}(\omega) \\ P^{(k)}_{qg}(\omega) & P^{(k)}_{gg}(\omega)
\end{array}\right)U_\omega
=\left(\begin{array}{ll}
P^{(k)}_{--}(\omega) &  P^{(k)}_{+-}(\omega) \\  P^{(k)}_{-+}(\omega) & P^{(k)}_{++}(\omega)
\end{array}\right),~~ (\mbox{hereafter } k\geq 1) \, , 
\label{matrix0}
%\end{equation}
\eea
where
%here $(k\geq1)$ and
the elements of  diagonalization martix $U_\omega$ are combinations of the LO anomalous dimensions, i.e.
\begin{equation}
U_\omega=\left(\begin{array}{ll}
1 & -1 \\ \frac{1-\alpha^{(0)}_\omega}{\varepsilon^{(0)}_\omega} & \frac{\alpha^{(0)}_\omega}{\varepsilon^{(0)}_\omega}
\end{array}\right),~~
U^{-1}_\omega=\left(\begin{array}{ll}
\alpha^{(0)}_\omega &~~~~ \varepsilon^{(0)}_\omega \\ \alpha^{(0)}_\omega-1 &~~~~ \varepsilon^{(0)}_\omega
\end{array}\right),
\label{matrix}
\end{equation}
with
\begin{equation}
\alpha^{(0)}_\omega=\frac{P_{qq}^{(0)}(\omega)-P_{++}^{(0)}(\omega)}
{P_{--}^{(0)}(\omega)-P_{++}^{(0)}(\omega)},\qquad
\epsilon^{(0)}_\omega=\frac{P_{gq}^{(0)}(\omega)}
        {P_{--}^{(0)}(\omega)-P_{++}^{(0)}(\omega)}
        %,\qquad
%\beta^{(0)}_\omega=\frac{P_{qg}^{(0)}(\omega)}{P_{--}^{(0)}(\omega)-P_{++}^{(0)}(\omega)}.
\label{elements}
\end{equation}
and
\begin{equation}
%\alpha =\frac{P_{qq}-P_{++}}{P_{--}-P_{++}},~~
%\varepsilon=\frac{P_{gq}}{P_{--}-P_{++}},~~
P^{(0)}_{\pm\pm}(\omega) = \frac{1}{2}\left[P^{(0)}_{qq}(\omega)+P^{(0)}_{gg}(\omega) \pm
\sqrt{(P^{(0)}_{qq}(\omega)-P^{(0)}_{gg}(\omega))^2+4P^{(0)}_{qg}(\omega)P^{(0)}_{gq}(\omega)}\right].\qquad
\label{Ppm0}
\end{equation}

The components $P^{(k)}_{--}(\omega)$ $(k\geq 1)$ of the timelike-splitting-function matrix can be obtained as \cite{Buras:1979yt}
%that
\begin{eqnarray}
P^{(k)}_{--}(\omega) &=& \alpha^{(0)}_\omega  P^{(k)}_{qq}(\omega) 
+ \epsilon^{(0)}_\omega P^{(k)}_{qg}(\omega) + 
\beta^{(0)}_\omega  P^{(k)}_{gq}(\omega)
+ (1-\alpha^{(0)}_\omega)  P^{(k)}_{gg}(\omega), \nonumber \\
 P^{(k)}_{-+}(\omega) &=& P^{(k)}_{--}(\omega) - 
\left(P^{(k)}_{qq}(\omega) +
\frac{1-\alpha^{(0)}_\omega}{\epsilon^{(0)}_\omega}  P^{(k)}_{gq}(\omega) \right), \nonumber \\
P^{(k)}_{++}(\omega) &=&  P^{(k)}_{qq}(\omega) +  P^{(k)}_{gg}(\omega) 
- P^{(k)}_{--}(\omega),
\nonumber \\
 P^{(k)}_{+-}(\omega) &=& P^{(k)}_{++}(\omega) - \left(P^{(k)}_{qq}(\omega) -
\frac{\alpha^{(0)}_\omega}{\epsilon^{(0)}_\omega}  P^{(k)}_{gq}(\omega) \right)
= P^{(k)}_{gg}(\omega) - 
\left(P^{(k)}_{--}(\omega) - \frac{\alpha^{(0)}_\omega}{\epsilon^{(0)}_\omega}  
P^{(k)}_{gq}(\omega) \right),\quad
\label{changebasis}
\end{eqnarray}
where
\be
\beta^{(0)}_\omega=\frac{P_{qg}^{(0)}(\omega)}{P_{--}^{(0)}(\omega)-P_{++}^{(0)}(\omega)} = \frac{\alpha^{(0)}_\omega(1-\alpha^{(0)}_\omega)}{\epsilon^{(0)}_\omega}
\, .
\label{beta0}
\end{equation}

The corresponding $\pm$ components of the fragmentation functions have the following form 
\begin{equation}
\left(\begin{array}{l} D_-(\omega) \\ D_+(\omega) \end{array}\right)
=U^{-1}_\omega \left(\begin{array}{l} D_s(\omega) \\ D_g(\omega) \end{array}\right)
=\left(\begin{array}{l}(\alpha^{(0)}_\omega D_s(\omega)+\varepsilon^{(0)}_\omega D_g(\omega) \\
\alpha^{(0)}_\omega-1)D_s(\omega)+\varepsilon^{(0)}_\omega D_g(\omega) \end{array}\right).
\label{ap1.2}
\end{equation}
%}

As it was mentioned above, the considered diagonalization gives a possibility to analyze FFs themselves.
  When we study the multiplicities, we should take into account the small $x$ resummation, which corresponds to
  the resummation near $N=1$ in the momentum space. It is quite easy to perform this resummation for the diagonal elements
  $P^{(k)}_{\pm\pm}(\omega)$ but rather difficult in the case of the non-diagonal ones  $P^{(k)}_{\pm\mp}(\omega)$
  (see discussion is Ref. \cite{Bolzoni:2013rsa}).

\subsection{Direct diagonalization of parton  multiplicities}

In the important simplification of QCD, namely
  ${\mathcal N}=4$ SYM,
  %super Yang-Mills (SYM) theory,
  a diagonalization was performed \cite{Kotikov:2002ab,Bianchi:2013sta} at all orders of perturbation theory,
  where the corresponding matrix contains the respective combinations of anomalous dimensions. 
Technically, it corresponds to the replacement  $P^{(0)}_{a,b}(\omega) \to P_{a,b}(\omega,a_s)$ as in (\ref{PijN}).
  \footnote{Strictly speaking, such replacement is directly applicable in the polarized case, while in spin-averaged case $(a,b=q,g,\varphi)$, since
    the contributions from scalars should be added.}.
%  this basis is actually more natural than the $(g,s)$ basis
%because the diagonal splitting functions $P^{(k)}_{\pm\pm}(\omega)$ may there be
%expressed in all orders of perturbation theory as one universal function with
%shifted arguments \cite{Kotikov:2002ab}.}

Following to this case, in Ref. \cite{Kniehl:2017fix} a similar diagonalization was
  applied to the case of multiplicities. So, we
%We
solved Eq.~(\ref{apR}) for $\omega=0$ exactly by exploiting Eqs.~(\ref{NNLL}) and (\ref{Basic}).
To this end, we diagonalize the NNLL DGLAP evolution kernel as
\begin{equation}
U^{-1}\left(\begin{array}{ll}
P_{qq} & P_{gq} \\ P_{qg} & P_{gg}
\end{array}\right)U
=\left(\begin{array}{ll}
P_{--} & 0 \\ 0 & P_{++}
\end{array}\right),~~
U=\left(\begin{array}{ll}
1 & -1 \\ \frac{1-\alpha}{\varepsilon} & \frac{\alpha}{\varepsilon}
\end{array}\right),~~
U^{-1}=\left(\begin{array}{ll}
\alpha &~~ \varepsilon \\ \alpha-1 &~~ \varepsilon
\end{array}\right),
\label{matrix}
\end{equation}
where
\begin{equation}
\alpha =\frac{P_{qq}-P_{++}}{P_{--}-P_{++}},~~
\varepsilon=\frac{P_{gq}}{P_{--}-P_{++}},~~
P_{\pm\pm} = \frac{1}{2}\left[P_{qq}+P_{gg}\pm
\sqrt{(P_{qq}-P_{gg})^2+4P_{qg}P_{gq}}\right] \, ,
\label{Ppm}
\end{equation}
where $P_{a,b}$ are given in Eqs. (\ref{NNLL}) and (\ref{nllfirstA}).

Acting by the operator $U^{-1}$ to Eq.~(\ref{apR}) for $N=1$ from the left (i.e. multiplying by the respective matrix) we rewrite it as
%thus assumes the form
\begin{equation}
U^{-1} \frac{\mu^2d}{d\mu^2}\left[ U U^{-1} \left(\begin{array}{l} D_s \\ D_g \end{array}\right)\right]
= U^{-1} \left(\begin{array}{ll}
P_{qq} & P_{gq} \\ P_{qg} & P_{gg}
\end{array}\right)U U^{-1} \left(\begin{array}{l} D_s \\ D_g \end{array}\right) \, .
\label{apR.1}
\end{equation}

Using the $\pm$ components of multiplicities as
\begin{equation}
\left(\begin{array}{l} D_- \\ D_+ \end{array}\right)
=U^{-1}\left(\begin{array}{l} D_s \\ D_g \end{array}\right)
=\left(\begin{array}{l}(\alpha D_s+\varepsilon D_g \\
\alpha-1)D_s+\varepsilon D_g \end{array}\right)\, ,
\label{ap1.2}
\end{equation}
we rewrite the above Eq. (\ref{apR.1}) as,
\begin{equation}
U^{-1} \frac{\mu^2d}{d\mu^2} \left[ U \left(\begin{array}{l} D_- \\ D_+ \end{array}\right)\right]
=
\left(\begin{array}{ll} P_{--} & 0 \\ 0 & P_{++}\end{array}\right)
%-U^{-1}\frac{\mu^2d}{d\mu^2}U\right]
\left(\begin{array}{l} D_- \\ D_+ \end{array}\right),
\label{apR.2}
\end{equation}

The essential difference between QCD and  ${\mathcal N}=4$ SYM is the $\mu^2$-dependence of the strong coupling constant, which propagates in turn to the matrix $U$. So, now the
matrix $U$ and the operator $\mu^2\frac{d}{d \mu^2}$ do not commute. To have a usual form of diagonal DGLAP equations, we can rewrite
Eq. (\ref{apR.2}) in the form
\begin{equation}
\mu^2 \frac{d}{d\mu^2}\left(\begin{array}{l} D_- \\ D_+ \end{array}\right)
=\left[
\left(\begin{array}{ll} P_{--} & 0 \\ 0 & P_{++}\end{array}\right)
-U^{-1}(\mu^2 \frac{d}{d\mu^2}) U\right]
\left(\begin{array}{l} D_- \\ D_+ \end{array}\right),
\label{ap2a}
\end{equation}
where the second term contained within the square brackets stems from the
commutator of $\mu^2\frac{d}{d \mu^2}$ and $U$
%, and

Owing to Eq.~(\ref{Basic}), the square root in
Eq.~(\ref{Ppm})
is exactly cancelled,
and we have
simple expressions for $P_{\pm \pm}$
\begin{eqnarray}
P_{--} = -A,~~ P_{++}=P_{qq}+P_{gg}+A,~~
\alpha = \frac{P_{gg}+A}{P_{qq}+P_{gg}+2A},~~
\varepsilon = -C \alpha \, .
\label{alpha1}
\end{eqnarray}

Let us stress, as it was already mentioned at the end of the previous section, that the presence of resummation of
anomalous dimensions and direct diagonalization of parton multiplicities leads to a 
%strong simplification of the values ​​of diagonal anomalous dimensions, in which all
strong simplification of the values of diagonal anomalous dimensions, in which all
 square roots have shrunk, which so far has only been observed in supersymmetric generalizations of QCD.

Inserting
Eq.~(\ref{alpha1}) in Eq.~(\ref{matrix}), we
have
\begin{equation}
U^{-1}(\mu^2\frac{d}{d\mu^2})U=-\frac{1}{\alpha}\,(\mu^2 \frac{d}{d\mu^2})\alpha
\left(\begin{array}{ll} 1 & 0 \\ 1 & 0\end{array}\right).
\label{AddTerm}
\end{equation}
Using the QCD $\beta$ function,
\begin{equation}
\mu^2 \frac{d}{d\mu^2} a_s=\beta(a_s)=-\beta_0a_s^2-\beta_1a_s^3+\mathcal{O}(a_s^4),~~
\beta_0=\frac{C_A}{3}(11-2\varphi),~~
\beta_1=\frac{2C_A^2}{3}[17-\varphi(5+3C)],
\label{eq:beta}
\end{equation}
after a small algebra
we may cast Eq.~(\ref{apR}) in its final form,
\begin{equation}
\mu^2 \frac{d}{d\mu^2}
\left(\begin{array}{l} D_- \\ D_+ \end{array}\right)
= \left(\begin{array}{ll} \frac{C\varphi\beta_0}{3C_A}\gamma_0^3-A &~~ 0 \\
\frac{C\varphi\beta_0}{3C_A}\gamma_0^3 &~~ P_{gg}+P_{qq} + A \end{array}\right)
\left(\begin{array}{l} D_- \\ D_+ \end{array}\right).\quad
\label{ap2b}
\end{equation}
The initial conditions are given by Eq.~(\ref{ap1.2}) for $\mu=\mu_0$ in terms
of the three constants $\alpha_s(\mu_0^2)$, $D_s(\mu_0^2)$, and $D_g(\mu_0^2)$.

\subsection{Results}

As seen from the Eq. (\ref{ap2b}), the
%The
"$-$" component $D_-$ can be obtained as
the general solution of a homogeneous differential equation. It has the following form \cite{Kniehl:2017fix}
\begin{equation}
\frac{D_-(\mu^2)}{D_-(\mu_0^2)} =
\exp\!{\left[\int_{\mu_0^2}^{\mu^2}\!\!\!\frac{d\bar{\mu}^2}{\bar{\mu}^2}
\!\left(\frac{C\varphi\beta_0}{3C_A}\gamma_0^3(\bar{\mu}^2)-A (\bar{\mu}^2) \!\right)\!\right]}
= \frac{T_-(\gamma_0(\mu^2))}{T_-(\gamma_0(\mu_0^2))}, ~~
\label{gensol.-}
\end{equation}
where
\begin{equation}
T_-(\gamma_0)=
\gamma_0^{d_-}\exp{\left(-\frac{4}{3}C\varphi \gamma_0\right)},~~ d_-=\frac{8C_A}{3\beta_0} \, C\varphi
\, .
\label{gensol.-T}
\end{equation}
%and
%%Here
%$\beta_0$ and $\beta_1$ are the first two terms of QCD $\beta$-function:
%\be
%\beta_0=\frac{C_A}{3}[11-2\varphi],~~
%\beta_1=\frac{2C_A^2}{3}[17-\varphi(5+3C)] \, .
%\label{beta}
%\end{equation}

The "$+$" component $D_+$ obeys \cite{Kniehl:2017fix}
to the inhomogeneous differential equation.
The general solution $\tilde{D}_+$ of its homogeneous part is
%reads
\bea
\frac{\tilde{D}_+(\mu^2)}{\tilde{D}_+(\mu_0^2)}=
&=&
\exp{\left[\int_{\mu_0^2}^{\mu^2}\frac{d\bar{\mu}^2}{\bar{\mu}^2}
\gamma_0(\bar{\mu}^2)\Bigl(1+ (2K_{q}^{(1)}+K_{g}^{(1)})
%K_+^{(1)}
\gamma_0(\bar{\mu}^2)+(K_{q}^{(2)}+K_{g}^{(2)})
%K_+^{(2)}
\gamma_0^2(\bar{\mu}^2)\Bigr)\right]}
\nonumber\\&=&
\frac{T_+(\gamma_0(\mu^2))}{T_+(\gamma_0(\mu_0^2))} \, ,
\label{gensol.+}
\eea
where
%, \nonumber \\&&
\be
T_+(\gamma_0)
=\gamma_0^{d_+}\exp{\left[\frac{4C_A}{\beta_0\gamma_0}
-\frac{4C_A}{\beta_0} \left(K_{q}^{(2)}+K_{g}^{(2)}
%K_+^{(2)}
-b_1\right)\gamma_0\right]},
%~~\beta_1=\frac{2C_A^2}{3}[17-\varphi(5+3C)],
~~ d_+=-\frac{4C_A}{\beta_0} \, (2K_{q}^{(1)}+K_{g}^{(1)})
\label{gensol.+T}
\ee
and $b_1=\beta_1/(2C_A \beta_0)$.

Adding to $\tilde{D}_+$ a special solution of the inhomogeneous differential
equation for $D_+$, we find its general solution \cite{Kniehl:2017fix}:
%to be
\begin{eqnarray}
D_+(\mu^2)=
%&=&
\left[\frac{D_+(\mu_0^2)}{T_+(\gamma_0(\mu_0^2))}
-\frac{4}{3}C\varphi\frac{D_-(\mu_0^2)}{T_-(\gamma_0(\mu_0^2))}
\int_{\gamma_0(\mu_0^2)}^{\gamma_0(\mu^2)}
\frac{d\overline{\gamma}_0}{1+b_1\overline{\gamma}_0^2}\,\frac{T_-(\overline{\gamma}_0)}{T_+(\overline{\gamma}_0)}\right]
T_+(\gamma_0(\mu^2)).
%\nonumber
\label{gensol.+a}
\end{eqnarray}

%\skip 0.5cm
%Using Eqs.~(\ref{matrix}) and (\ref{ap1.2}), we

\section{Casimir scaling}

In the supersymmetric generalization of QCD (SQCD)
%SQCD
the corresponding relation (i.e. (\ref{Basic}) with $C=1$) exists
\cite{Dokshitzer:1977sg,Kom:2012hd,Kounnas:1982de} for
the anomalous dimensions $P_{ab}^{SUSY}(N)$ with arbitrary $N$ values \footnote{In fact it was observed for the splitting functions $P_{ab}^{SUSY}(x)$, which correspond to the $P_{ab}^{SUSY}(N)$ (see Eq. (\ref{PabN})).}:
%in the Bjorken $x$ space: $P_{ab}^{SUSY}(N)=\int^1_0 \, dx^{N-1} P_{ab}^{SUSY}(x)$.}:
\be
P_{qq}^{SUSY}(x) + P_{gq}^{SUSY}(x) = P_{gg}^{SUSY}(x) + P_{qg}^{SUSY}(x)\, .
\label{BasicSUSY}
\ee

Beyond LO the
%The
property (\ref{BasicSUSY}) is violated
%beyond LO
in the standard "dimension regularization"
but it survives in the form of the "dimensional reduction" \cite{Siegel:1979wq}
%, which
and was
%and may be
used
also to check real calculations (see Refs. \cite{Mertig:1995ny,Antoniadis:1981zv} and discussion therein).
It seems that the relation (\ref{BasicSUSY}) is violated
\cite{Almasy:2011eq} at the NNLO level of accuracy but this requires some additional investigations.

It will be interesting to check whether Eq.~(\ref{Basic}) also holds beyond
$\mathcal{O}(\gamma_0^3)$ in the case of
the "dimensional reduction" \cite{Siegel:1979wq}.
The choice of a scheme in the
%present
above consideration
%at the accuracy $\mathcal{O}(\gamma_0^3)$
is not  so important because a difference in the results of various schemes
  is exactly cancelled in Eq. (\ref{Basic}).

Following to \cite{Dokshitzer:2008zz}, Eq. (\ref{BasicSUSY}) can be spelled out as an equality of the total probabilities of "quark" and "gluon" decays.
We note that such probabilistic interpretation becomes to be very important directly in QCD
  %(see Ref.
  \cite{Teryaev:1998iw,Artru:2008cp}
  %)
for decoupling of orbital and total angular momenta in nucleon.

  Following to \cite{Dokshitzer:2008zz,Teryaev:1998iw},
we can explore the probabilistic properties hidden in Eq. (\ref{Basic}).
To do it, we introduce
new form of the quark  $\overline{D}_s$ and gluon $\overline{D}_g$ multiplicities
\be
D_s(\mu^2) = C_F \overline{D}_s(\mu^2),~~ D_g(\mu^2) = C_A \overline{D}_g(\mu^2) \, ,
\label{NewPDF}
\ee
where we extract the corresponding "color charges" $C_F$ and $C_A$, respectively.

The new multiplicities obey to the following DGLAP equations
\be
%\frac{\mu^2d}{d\mu^2} \, \overline{D}_a(\mu^2) = \sum_{b=s,g} \overline{P}_{ba}(\alpha_s(\mu^2))  \overline{D}_b(\mu^2) \, ,
\mu^2 \frac{d}{d\mu^2}
\left(\begin{array}{l} \overline{D}_s(\mu^2) \\ \overline{D}_g(\mu^2) \end{array}\right)
=\left(\begin{array}{ll} \overline{P}_{qq} & \overline{P}_{gq} \\
\overline{P}_{qg} & \overline{P}_{gg} \end{array}\right)
\left(\begin{array}{l} \overline{D}_s(\mu^2) \\ \overline{D}_g(\mu^2) \end{array}\right),
\label{DGLAPnew}
\ee
where
\be
\overline{P}_{aa} = P_{aa},~~
%~~(a=s,g),~~
\overline{P}_{qg} = C \, P_{qg},~~ \overline{P}_{gq} = C^{-1} \, P_{qq} \,
\label{ADnew}
\ee
and the relation (\ref{Basic}) becomes to be as follows
\begin{equation}
\overline{P}_{qq}+ \overline{P}_{gq} = \overline{P}_{gg} + \overline{P}_{qg} \, ,
\label{BasicNew}
\end{equation}
i.e. it exactly equals (for $N=1$) to the one in (\ref{BasicSUSY}) obtained in the SQCD framework.
%of SUSY QCD.

So, the new parton multiplicities $\overline{D}_a$ have the same probabilistic properties as the
original ones $D_a$
%interpretation
in the supersymmetric case bringing it closer to
%the observed values
%bringnging SUSY closer to
observable quantities. 

Since the parton multiplicities $\overline{D}_a$ are proportional to the standard ones  $D_a$,
the solution of the DGLAP equation (\ref{DGLAPnew}) is the same as one done in Ref. \cite{Kniehl:2017fix} for the equation (\ref{apR}) at $N=1$: after diagonalization of (\ref{DGLAPnew}) there are two solutions in the form of so-called "$+$" and "$-$" components.

\subsection{High-energy asymptotics of multiplicities}
Returning
%We now return
to the parton basis,
%where
it is useful to decompose $\overline{D}_a=\overline{D}_a^++\overline{D}_a^-$ into the large and small components $\overline{D}_a^\pm$ proportional to $D_\pm$, respectively.
Defining $\overline{r}_\pm=\overline{D}_g^\pm/\overline{D}_s^\pm$ and using Eqs.~(\ref{NNLL}), (\ref{nllfirstA}),
and (\ref{gensol.+}), we then have $C_F \overline{D}_s^\pm=\mp D_\pm$ and
\begin{equation}
\overline{r}_+ =
%-\frac{\alpha}{\epsilon}=\frac{1}{C}
1+\mathcal{O}(\gamma_0^2),~~
\overline{r}_-=
%\frac{1-\alpha}{\epsilon} =
-\frac{4}{3} n_f
%\varphi
\gamma_0
+\frac{n_f}{18}[29-2\varphi(5-2C)]\gamma_0^2
+\mathcal{O}(\gamma_0^3).
\label{eq:rpm}
\end{equation}
Recalling that $\overline{r}=\overline{D}_g/\overline{D}_s$,
%$\langle n_h\rangle_q=D_s$ and $\langle n_h\rangle_g=D_g$,
we
%thus
have
\begin{equation}
\overline{r}= \frac{\overline{r}_+ + \overline{r}_-\overline{D}_s^-/\overline{D}_s^+}{1+
\overline{D}_s^-/\overline{D}_s^+}
%\bigl(r_++r_-D_s^-/D_s^+\bigr)/\bigl(1+D_s^-/D_s^+\bigr).
  \label{eq:r}
\end{equation}

So, for the high energy asymptotics (i.e. $\mu \to \infty$), where the "$+$"-component
strongly dominates, we have for the ratio $\overline{r}$:
\be
\overline{r} \to \overline{r}_+ =1 \, ,
\label{High}
\ee
i.e. the new multiplicities of gluon and quark jets become to be equal in all known orders.
This equality
  %(\cite{High})
  corresponds exactly to the  
    Casimir scaling (i.e. to $D_g^+/D_s^+ = C_A/C_F$) mentioned above.
    One shoud expect that  
      %Moreover, this equality  should be clearly seen experimantally and thus can be considered as a bechmark in high-energy experiments.}
      %Moreover,
      this equality should be clearly seen experimentally and thus can be regarded as a guideline in high energy experiments and a complementary tool 
     for discrimination of quark and gluon jets. 

    When going to lower energy values, this equivalence should
    % may
    be violated. One of the important elements of the violation is the appearance of contributions proportional to the quartic Casimirs in
    high orders of the perturbation theory (see
    %similar
    investigations in Refs. \cite{Anzai:2010td,Catani:2019rvy,Lee:2016cgz,Henn:2019rmi,Das:2019btv}).
  So, we think that the equality $\overline{r}_+ =1$
%  (\ref{High})
  may
%will
  be kept  up to $a_s^4 \sim \gamma_0^8$ accuracy\footnote{Note that the quartic Casimir contributions may be negligible numerically
      %\cite{Bolzoni:2012ii}
  \cite{Catani:2019rvy} and the ``Casimir scaling'' may be fulfilled even above $a_s^4 \sim \gamma_0^8$ accuracy in approximated
  form.}, where the corresponding  splitting functions
  $P_{ba}$ would contain Feynman diagrams coming with the quartic Casimir contributions.
  %, as it was investigated recently
%  (see similar investigations in Refs. \cite{Anzai:2010td,Catani:2019rvy,Lee:2016cgz,Henn:2019rmi,Das:2019btv}).
%  in all orders of perturbation theory.

  However, this is not the only source of violation of the property $\overline{r} =1$. As seen from the eq. (\ref{eq:r}),
the existence of the "$-$"-component violates the equality (\ref{High})
between the new multiplicities
$\overline{D}_s$ and $\overline{D}_g$
%already at the accuracy $O(\gamma_0)$
that may be
%is
essentially stronger than the 
possible violation due the quartic Casimir contributions.
Of course, the  "$-$"-component does not increase with energy increasing, but its contribution leads to a nontrivial
  dependence of the ratio of gluon and quark multiplicities, which is important at non-asymptotically large energy values.
  Such nontrivial dependence is seen in experimental data (see Refs. \cite{Kniehl:2017fix,Bolzoni:2013rsa,Bolzoni:2012ii} and
  discussions therein).
We note that the ratio $\overline{r}_- \sim n_f$ and thus the equality (\ref{High})
should be violated in pure gluodynamics essentially slowly, i.e. at  $a_s^4 \sim \gamma_0^8$ accuracy by contributions of
%  due
the quartic Casimirs.
% contributions.}

 We note also that the contribution of  the "$-$"-component is very important
\cite{Kniehl:2017fix,Bolzoni:2013rsa,Bolzoni:2012ii}
 for comparison of the theoretical predictions for the parton jet multiplicities with the experimental data, which belongs
 to the subasymptotic range. Indeed, as it was shown in \cite{Bolzoni:2013rsa,Bolzoni:2012ii}, the
   % The
   "$-$"-component contribution gives the natural explanation of the longstanding discrepancy in the theoretical description of
   the data, which was reviewed, for example, in Ref. \cite{Dremin:2000ep}.
 In a sense, the presence of the  "$-$"-component leads to a rather different evolution of the gluon and quark multiplicities, which
   is in good agreement with experimental data.

The importance of the  "$-$"-component contribution
  %It
  is also in full agreement with the study \cite{Kotikov:1998qt} of low $x$ asymptotics of
 parton densities, where the existence of the corresponding "$-$"-component leads to a good
 %is strongly important to have
 agreement between theoretical studies \cite{Kotikov:2012sm}
 and the experimental data \cite{Aaron:2009aa}
 for the structure function $F_2(x,Q^2)$
 of the deep-inelastic scattering obtained by H1 and ZEUS Collaborations.

\section{Conclusions}
 
In this article,
%summary,
we study in some detail
the SUSY-like relation \cite{Kniehl:2017fix} between the
NNLL-resummed first Mellin moments of the timelike DGLAP splitting functions
in real QCD, in Eq.~(\ref{Basic}).
This relation appeared through the small-$x$ resummation of the time-like splitting functions
  and non-standard diagonalization \cite{Kniehl:2017fix} of their first Mellin moments. In sections 2 and 3 we presented
  the basic steps of the
  %small-$x$
  resummation
  %of the time-like splitting functions
  and the
  %nonusual
  diagonalization, respectively.
  %, which are responsable for
  %the evolution of parton multiplicities.}

In Eq. (\ref{NewPDF}) we
%We
introduced the new quark and gluon jet multiplicities $\overline{D}_a$
%$\overline{D}_s$ and $\overline{D}_g$,
which have probabilistic properties,
%which are
same as for the standard multiplicities $D_a$
in the framework of the supersymmetric extension of QCD.
 %As shown above,
  As it was already discussed in Section 4.1, the
  %The
  new quark and gluon multiplicities $\overline{D}_a$ should have similar behavior at high energies that
   can be regarded as a guideline in high energy experiments.
  The
  %nature of the
  violation of this similarity in the region of lower energies is controlled by the
  violation of the Casimir scaling, both due to the appearance of contributions proportional to the quartic Casimirs,
  and due to the "-" component. At high energies, these effects are small, which is associated with the fact that
  the contributions proportional to the quartic Casimirs are suppressed by $\sim \alpha_s^2$, and the "-" component
  does not contain the factor $\sim \exp[\sim 1/\sqrt{\alpha_s}]$ growing in the high-energy region, and thus is
  also strongly suppressed.
With decreasing energy, these effects will begin to manifest themselves, leading to different energy dependences of
  quark and gluon multiplicities.

Such a similarity of parton multiplicities $\overline{D}_a$ at high energies and the appearance of differences at lower energies
can be studied in experiments performed at LHC and at future Electron-Ion Collider. 

\section{Acknowledgments}

A.V.K. thanks Prof. Bernd Kniehl for the joint work during the preparation of Ref. \cite{Kniehl:2017fix},
which is the base for current investigations.
The work of O.V.T. was partially supported by RFBR grant 18-02-01107.

%{\bf Acknowledgements}\\

\end{document}